\documentclass[prl,twocolumn,amssymb,showpacs,amsmath,nobibnotes,floatfix,aps]{revtex4}
\usepackage[]{graphicx,axodraw,bm}

\newcommand{\lsim}{\lesssim}
\renewcommand{\Re}{\ensuremath{\mathop{\rm Re}}}
\newcommand{\arrow}{\!\to\!}

\begin{document}
\noindent {\scriptsize \hspace*{-0.011cm} FERMILAB-Pub-08-051-T}
\title{\boldmath Accumulating
evidence for nonstandard leptonic decays of $D_s$ mesons}
\author{Bogdan A. Dobrescu and Andreas~S.~Kronfeld}
\affiliation{Theoretical Physics Department,
Fermi National Accelerator Laboratory, Batavia, Illinois, USA}

\date{March 4, 2008; revised April 28, 2008}
\begin{abstract}
The measured rate for $D_s^+\arrow \ell^+\nu$ decays, where $\ell$ 
is a muon or tau, is larger than the standard 
model prediction, which relies on lattice QCD, at the 3.8$\sigma$ level. 
We discuss how robust the theoretical prediction is, and 
we show that the discrepancy with experiment may be explained 
by a charged Higgs boson or a leptoquark.
\end{abstract}

\maketitle

%%%%%%%%%%%%%%%%%%%%%%%%%%%%%%%%%%%%%%%%%%%%%%%%%

{\it Introduction.}---The pattern of flavor and $CP$ violation of the
standard model has been established by a wide range of experiments.
This agreement, however, leaves room for new flavor effects to show up
as calculations and measurements improve.
Intriguingly, decays of the $D_s$ meson 
(the lightest $c\bar{s}$ state) could be more sensitive to new
physics than any other process explored so far.
It suffices that a new particle couples predominantly to leptons and
up-type quarks, but not to the first generation.
%those of the first generation.

In this Letter we examine the leptonic decays of the~$D_s$.
Recently, the calculation of the relevant QCD matrix element
has improved significantly, and more accurate measurements of the rate
have been made. 
The average of the experimental results disagrees with the standard
model by almost four standard deviations.
We discuss the evidence, and  
%several explanations, including the possibility that 
propose that
a nonstandard amplitude interferes with the standard
$W$-mediated amplitude.
We show that the tree-level exchange of a spin-0 particle with mass 
of order 1 TeV may account for the discrepancy.

\medskip
%%%%%%%%%%%%%%%%%%%%%%%%%%%%%%%%%%%%%%%%%
{\it Leptonic $D_s$ decays.}---The  $D_s\arrow\ell\nu$ branching fraction,
where $\ell$ is a charged lepton of mass
$m_\ell$, is given in the standard model by
\begin{equation}
\hspace*{-0.8em}	B(D_s\arrow\ell\nu) = \frac{m_{D_s}}{8\pi} \tau_{D_s}f_{D_s}^2
		\left| G_F V^*_{cs}m_\ell\right|^2 
		\left(1\!-\frac{m_\ell^2}{m_{D_s}^2}\right)^{\!\! 2}.
	\label{eq:rate}
\end{equation}
Here $m_{D_s}$ and $\tau_{D_s}$ are the mass and lifetime of the~$D_s$, 
$G_F$~is the Fermi constant, 
and $V_{cs}$ is a Cabibbo-Kobayashi-Maskawa (CKM) element.
The decay constant $f_{D_s}$ is defined by
\begin{equation}
	\langle0| \, \bar{s}\gamma_\mu\gamma_5 c \, |D_s(p)\rangle = 
		if_{D_s}p_\mu,
	\label{eq:0AD}
\end{equation}
where $p_\mu$ is the 4-momentum of the $D_s$ meson.
Although the electroweak transition proceeds at the tree level,
$D_s^+\arrow W^+\arrow\ell^+\nu_\ell$, its rate is suppressed.
The helicity of the lepton must flip, leading to the factor $m_\ell$ in 
the amplitude. For the muon, this helicity suppression $(m_\ell/m_{D_s})^2$ 
is $2.8\times10^{-3}$.
The $\tau$ mass is only 10\% smaller than the $D_s$ mass (1.969 GeV),
so there is no significant helicity suppression, but the phase space
suppression [the last factor in Eq.~(\ref{eq:rate})] is
$3.4\times10^{-2}$.

\begin{table}[tp]
	\centering
	\caption{Experimental values of $f_{D_s}$.  
	Our averages treat systematic uncertainties as uncorrelated and 
	omit the PDG entry~\cite{Yao:2006px}, which is an
	average of earlier experiments.}
	\label{tab:fDs}
	\begin{tabular}{l@{\quad}l@{\quad}l}
		\hline\hline
		final state & reference  & $f_{D_s}$~(MeV)  \\
		\hline
		$\ell\nu$ & PDG~\cite{Yao:2006px}       & $294\pm27$ \\
		$\mu\nu$  & BaBar~\cite{Aubert:2006sd}  & $283\pm17\pm16$  \\
		$\mu\nu$  & CLEO~\cite{Pedlar:2007za}   & $264\pm15\pm\;7$ \\
		$\mu\nu$  & Belle~\cite{Widhalm:2007mi} & $275\pm16\pm12$  \\
		$\tau\nu$ ($\tau\arrow\pi\nu$) & 
			CLEO~\cite{Pedlar:2007za}   & $310\pm25\pm~8$ \\
		$\tau\nu$ ($\tau\arrow e\nu\bar{\nu}$) \ \ \ & 
			CLEO~\cite{Ecklund:2007zm}  & $273\pm16\pm\;8$ \\
		$\mu\nu$  & our average \ \ \ & $273\pm11$  \\
		$\tau\nu$ & our average & $285\pm15$ \\
		\hline\hline
	\end{tabular}
\end{table}

We have collected in Table~\ref{tab:fDs} all precise experimental 
measurements of  $B(D_s\arrow\ell\nu)$, which are usually quoted 
in terms of $f_{D_s}$~\cite{Yao:2006px,Rosner:2008yu}.
Combining the error bars in quadrature, our average of $\tau\nu$ and
$\mu\nu$ final states is
\begin{equation}
	\left(f_{D_s}\right)_{\rm expt} = 277 \pm 9 \; \textrm{MeV} .
	\label{eq:fDs-expt}
\end{equation}
The most accurate calculation from lattice QCD is~\cite{Follana:2007uv}
\begin{equation}
	\left(f_{D_s}\right)_{\rm QCD} = 241 \pm  3 \; \textrm{MeV} ,
	\label{eq:fDs-HPQCD} 
\end{equation}
where statistical and systematic uncertainties are combined in the 
fitting methods.
The only other modern lattice-QCD calculation agrees,
$249\pm3\pm16$~MeV \cite{Aubin:2005ar}, but its quoted error is
five times larger and would not influence a weighted average with
Eq.~(\ref{eq:fDs-HPQCD}).
The discrepancy between Eqs.~(\ref{eq:fDs-expt}) and (\ref{eq:fDs-HPQCD}) 
is 15\% and 3.8$\sigma$.
Table~\ref{tab:fDs} also shows averages for each mode separately: for 
$\tau\nu$ ($\mu\nu$) alone, the discrepancy is 18\% and 2.9$\sigma$
(13\% and 2.7$\sigma$).

If the BaBar result is omitted from the average, 
as in Ref.~\cite{Rosner:2008yu}, 
then the discrepancy is 3.4$\sigma$.
On the other hand, if the earlier measurements \cite{Yao:2006px} 
as well as the BaBar result are included, 
we find a 4.1$\sigma$ discrepancy. 

\medskip
%%%%%%%%%%%%%%%%%%%%%%%%%%%%%%%%%%%%%%%%%
{\it Experiments.}---CLEO~\cite{Pedlar:2007za,Ecklund:2007zm} 
produces $D_s$ pairs near threshold, where the multiplicity is low.
Their method reconstructs one $D_s^{(*)}$ and then counts 
how often the opposite-side $D_s$ decays leptonically.
When the charged lepton is a muon, the neutrino is ``detected'' by
requiring the missing mass-squared to peak at zero. 
When the charged lepton is a $\tau$, the identification is made through 
the subsequent decays $\tau\arrow e\nu\bar{\nu}$ and 
$\tau\arrow\pi\bar{\nu}$.
BaBar~\cite{Aubert:2006sd} observes $D_s$ coming from the decay
$D_s^*\arrow D_s\gamma$, produced well above threshold.
They compare the relative number of subsequent $D_s\arrow\mu^+\nu$ and 
$D_s\arrow\phi\pi$, and then use their own measurement of
$B(D_s\arrow\phi\pi)$ %\cite{Aubert:2005xu} 
to determine $B(D_s\arrow\ell\nu)$.
Belle~\cite{Widhalm:2007mi} also observes $D_s$ via 
$D_s^*\arrow D_s\gamma$, but the whole event is reconstructed, 
using a Monte Carlo technique.
In summary, all these measurements have central values and error bars 
that are straightforward to interpret, and to combine to obtain 
Eq.~(\ref{eq:fDs-expt}).

The measured branching fraction and Eq.~(\ref{eq:rate})
yield $|V_{cs}|f_{D_s}$.
Three-generation CKM unitarity is assumed, either taking
$|V_{cs}|$ from a global fit to flavor physics~\cite{Yao:2006px}, or
setting $|V_{cs}|=|V_{ud}|$.
The difference is numerically irrelevant.
Relaxing the assumption cannot lead to agreement between theory and
experiment because unitarity, even for more than three generations,
requires $|V_{cs}|< 1$, whereas the discrepancy would require
$|V_{cs}|\approx 1.1$.

\medskip
%%%%%%%%%%%%%%%%%%%%%%%%%%%%%%%%%%%%%%%%%
{\it Radiative corrections.}---The measurements are not, strictly 
speaking, for $D_s\arrow\ell\nu$ alone, because some photons 
are always radiated.
The radiative corrections have been studied, focusing on effects that 
could overcome the helicity suppression~\cite{Burdman:1994ip,Hwang:2005uk}.

For $D_s\arrow\tau^+\nu$ there is no sizable helicity suppression.
In the rest frame of the $D_s$, the $\tau$ acquires only 9.3~MeV of 
kinetic energy, so it cannot radiate much.
Explicit calculation~\cite{Burdman:1994ip} shows that the radiative
corrections are too small to account for the
discrepancy~\cite{Wang:2001mm}.

For $D_s\arrow\mu^+\nu$ radiative corrections could play a role due to 
processes of the form 
$D_s\arrow\gamma D_s^*\arrow\gamma\mu^+\nu$, where $D_s^*$ is a (virtual) 
vector or axial-vector meson.
The transition $D_s^*\arrow\mu^+\nu$ is not helicity-suppressed, so the
factor $\alpha$ for radiation is compensated by a relative factor
$m_{D_s}^2/m_\mu^2$ for omitting helicity suppression.
Using Eq.~(12) of Ref.~\cite{Burdman:1994ip} and imposing the 
CLEO~\cite{Pedlar:2007za} cut $E_\gamma>300$~MeV, we find that the
radiative rate is around 1\% and, hence, insufficient to
explain the discrepancy.

\medskip
%%%%%%%%%%%%%%%%%%%%%%%%%%%%%%%%%%%%%%%%%
{\it Lattice QCD}---There are many lattice-QCD calculations for $f_{D_s}$
in the literature, but only
Refs.~\cite{Follana:2007uv,Aubin:2005ar} include 2+1 flavors of
sea quarks, which is necessary to find agreement for many
``gold-plated'' quantities, namely those for which errors are easiest 
to control~\cite{Davies:2003ik}.
Both calculations start with lattice gauge fields generated 
by the MILC Collaboration~\cite{Bernard:2001av}, which 
employ ``rooted staggered fermions'' for the sea quarks.
At finite lattice spacing this approach has small violations of 
unitarity and locality.
Theoretical and numerical evidence suggests that these vanish in the
continuum limit, such that QCD is obtained, with the undesirable 
features controlled with chiral perturbation theory.
The strengths and weaknesses of this approach have been reviewed in 
detail~\cite{Sharpe:2006re}.

Reference~\cite{Follana:2007uv} reports an error five times smaller than 
that of Ref.~\cite{Aubin:2005ar} for several reasons.
The largest uncertainties in Ref.~\cite{Aubin:2005ar} come from a 
power-counting estimate of the discretization error for the charm 
quark, and from uncertainties in the chiral extrapolation.
Reference~\cite{Follana:2007uv} employs a different discretization for
the charm quark, which allows a controlled extrapolation to the
continuum limit.
Thus, the discretization error here is driven by the underlying 
numerical data.

The action for the charm quark in Ref.~\cite{Follana:2007uv}, called 
HISQ~\cite{Follana:2006rc}, is the same as that used for the light 
valence quarks.
As a result the statistical errors are smaller than those of the 
heavy-quark method used in Ref.~\cite{Aubin:2005ar}, and 
the axial current automatically has the physical normalization.
The suitability of HISQ for charm is one of its design features, it 
has been tested via the charmonium spectrum~\cite{Follana:2006rc},
and the computed $D$ and $D_s$ masses agree with experiment.
The $D^+$ decay constant~$f_{D^+}$ also agrees with experiment, at 
$1\sigma$.

Another feature of Ref.~\cite{Follana:2007uv} is the way the 
lattice-spacing and sea-quark mass dependence is fitted.
Full details are not yet published, but it is noteworthy that the same
analysis yields $f_\pi$ and $f_K$ in agreement with
experiment~\cite{Yao:2006px} and earlier, equally precise, lattice-QCD
calculations~\cite{Aubin:2004fs}.
The $D_s$ meson is simpler than the pion or kaon for lattice QCD,
because none of the valence quarks is light, so $f_{D_s}$ is 
easier to determine than $f_\pi$.
We find that simple extrapolations lead to the same central values for
both $m_{D_s}$ and~$f_{D_s}$.
 
% The result shown in Eq.~(\ref{eq:fDs-HPQCD}) appears, thus, to be 
% solid.
% Similarly precise computations of $f_{D_s}$ with other methods for
% lattice fermions should be carried out, but we do not see a simple 
% reason to expect a substantial shift.
% Furthermore, even if the error bar were doubled, the discrepancy would
% remain significant: 2.7$\sigma$, 2.5$\sigma$, and 3.3$\sigma$
% for $\tau$, $\mu$, and combined.
% This follows from the fact that the discrepancy is dominated by the 
% experimental uncertainties.

The error bar in Eq.~(\ref{eq:fDs-HPQCD}) is smaller than that in
Eq.~(\ref{eq:fDs-expt}).
Therefore, it is the combined experimental error that provides the
yardstick for the deviation.
To illustrate, if the lattice-QCD error bar were doubled, the
discrepancy becomes 2.7$\sigma$, 2.5$\sigma$, and 3.3$\sigma$ for
$\tau$, $\mu$, and combined.
Hence, even if additional sources of uncertainty are uncovered, 
evidence for a deviation may well remain.

\medskip
%%%%%%%%%%%%%%%%%%%%%%%%%%%%%%
{\it Nonstandard effective interactions.}---Although the experiments 
quote the final states as $\mu^+\nu_\mu$ and
$\tau^+\nu_\tau$ (and their charge conjugates), the flavor of the
neutrino is not detected.
Nonstandard physics could lead to any neutrino flavor, even a sterile
neutrino.
However, given the large effect that needs to be explained, we shall
restrict our attention to amplitudes that could interfere with the
standard model, which fixes the neutrino flavor.  
Lorentz-invariant new physics may contribute to $D_s\arrow\ell\nu_\ell$ 
only through the following effective Lagrangian: 
\begin{equation}	
	\frac{C_A^\ell}{M^2} \left( \bar{s} \gamma_\mu \gamma_5 c \right)
		\left(\bar{\nu}_L \gamma^\mu \ell_L \right) +
	\frac{C_P^\ell}{M^2} \left( \bar{s} \gamma_5 c \right)
		 \left(\bar{\nu}_L \ell_R \right) + {\rm H.c.},
	\label{eq:Leff4}
\end{equation}
where $C_A^\ell$ and $C_P^\ell$ are complex dimensionless parameters,
$M$ is the mass of some particle whose exchange induces the 4-fermion
operators~(\ref{eq:Leff4}), and the $c,s,\ell$ fields are taken in the
mass-eigenstate basis.
The hadronic matrix element required for the decay induced by
$(\bar{s}\gamma_5c)(\bar{\nu}_L\ell_R)$ is related to the one of
Eq.~(\ref{eq:0AD}) by partial conservation of the axial current:
%\begin{equation}
%	\langle 0| \, \bar{s}\gamma_5 c \, |D_s\rangle = -
%		if_{D_s}m^2_{D_s}(m_c+m_s)^{-1}.
%	\label{eq:0PD}
%\end{equation}
$(m_c+m_s)\langle 0| \, \bar{s}i\gamma_5 c \, |D_s\rangle =f_{D_s}m^2_{D_s}$.
The branching fraction in the presence of the operators (\ref{eq:Leff4}) 
is given by  Eq.~(\ref{eq:rate}) with $G_FV^*_{cs}m_\ell$ replaced by
\begin{equation}
	 G_F V^*_{cs} m_\ell + \frac{1}{\sqrt{2}M^2} 
\left(  C_A^\ell m_\ell + \frac{C_P^\ell \, m_{D_s}^2}{m_c+m_s}\right) ~,
	\label{eq:lq-rate}
\end{equation}
with no helicity suppression in the last term.

The imaginary part of $V_{cs}$ is negligible (in the standard CKM
parametrization~\cite{Yao:2006px}), so constructive interference,
which would increase $B(D_s\arrow\ell\nu)$,
requires the real part of $C_A^\ell$ or $C_P^\ell$ to be nonzero and
positive.
Assuming only one nonzero coefficient, the amplitude for $\tau^+\nu_\tau$ 
($\mu^+\nu_\mu$) could be increased by 12\% (8.4\%) only if
\begin{eqnarray}
	\frac{M}{(\Re C_A^\ell)^{1/2}}  & \lesssim & \left\{
	\begin{array}{rl}
		710~\textrm{GeV} & \textrm{for}~\ell=\tau \\  [1mm]
		\hphantom{3}850~\textrm{GeV} & \textrm{for}~\ell=\mu
	\end{array}
	\right. \; , \label{eq:estimation-A} \\  [1mm]
	\frac{M}{(\Re C_P^\ell)^{1/2}}  & \lesssim & \left\{
	\begin{array}{rl}
		920~\textrm{GeV} & \textrm{for}~\ell=\tau \\ [1mm]
		4500~\textrm{GeV} & \textrm{for}~\ell=\mu
	\end{array}
	\right.  \; , \label{eq:estimation-P}
\end{eqnarray}
thereby reducing the discrepancy to $1\sigma$ in each case.
These bounds are a key new result of this Letter, because they constrain
any model of new physics.

The effective interaction~(\ref{eq:Leff4}) also contributes to the 
semileptonic decays $D\arrow K\mu^+\nu$.
This proceeds through two amplitudes, corresponding to
angular momentum $J=1$ or $0$ for the lepton pair.
For $J=1$, the standard-model amplitude and that from $C_A^\mu$
are not helicity suppressed, while that from $C_P^\mu$ is.
For $J=0$, the pattern of helicity suppression is as for the leptonic decay.
% The tensor interactions are also helicity suppressed.
Hence, only the $J=1$ part of the rate will be visible, and 
as the accuracy of the lattice-QCD calculations improves, the 
comparison with experiment will help decide which interactions are 
responsible for the effect in $D_s\arrow\ell\nu$.
The current status favors $C_P^\mu\neq 0$ rather than $C_A^\mu\neq 0$, 
because the lattice-QCD prediction for $D\arrow K\mu\nu$~\cite{Aubin:2004ej} 
agrees with experiment~\cite{Widhalm:2006wz}, albeit at the $\sim7\%$
level.

\begin{figure}[b]
\begin{center}
\scalebox{0.81}{
\unitlength=1 pt
\SetScale{1}\SetWidth{1}      % line    size control
\normalsize\large  
{} \allowbreak
\begin{picture}(180,50)(51,20)
\SetWidth{1} 
\put(0,20){
\ArrowLine(20,20)(0,0)\ArrowLine(0,40)(20,20)
\ArrowLine(90,40)(70,20)\ArrowLine(70,20)(90,0)
\DashArrowLine(20,20)(70,20){4}
\Text(-9,37)[c]{$c$}\Text(-9,5)[c]{$\bar{s}$}
\Text(103,37)[c]{$\ell^+$}\Text(99,5)[c]{$\nu$}
\Text(45,30)[c]{$(+1)$}
}
\put(137,0){
\ArrowLine(20,20)(0,0)\ArrowLine(40,0)(20,20)
\ArrowLine(20,60)(40,80)\ArrowLine(0,80)(20,60)
\DashArrowLine(20,60)(20,20){3}
\Text(-5,70)[c]{$c$}\Text(-5,10)[c]{$\bar{s}$}
\Text(44,70)[c]{$\nu$}\Text(49,10)[c]{$\ell^+$}
\Text(44,40)[c]{$ (+2/3)$}
}
\put(220,0){
\ArrowLine(20,20)(0,0)\ArrowLine(20,20)(40,0)
\ArrowLine(40,80)(20,60)\ArrowLine(0,80)(20,60)
\DashArrowLine(20,60)(20,20){3}
\Text(-5,70)[c]{$c$}\Text(-5,10)[c]{$\bar{s}$}
\Text(46,70)[c]{$\ell^+$}\Text(45,10)[c]{$\nu$}
\Text(44,40)[c]{$(-1/3)$}
}
\end{picture}
}
\end{center}
\caption{Four-fermion operators induced by boson exchange.
}
\label{fig:diagrams}
\end{figure}
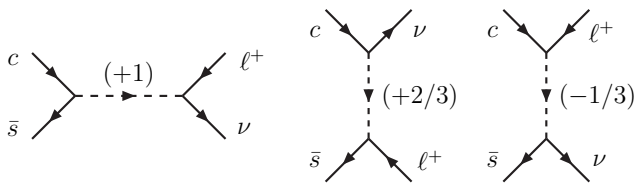

\medskip
%%%%%%%%%%%%%%%%%%%%%%%%%%%%%%
{\it New particles.}---There are three 
choices for the electric charge of a boson that can
mediate the four-fermion operators (\ref{eq:Leff4}): $+1, +2/3,
-1/3$, corresponding to the three diagrams shown in
Fig.~\ref{fig:diagrams}.
The exchanged boson (taken to be emitted from the vertex where $c$ is
absorbed) is a color singlet if the electric charge is +1, and a color
triplet if the electric charge is $+2/3$ or $-1/3$.
We shall consider only the cases where the new boson has spin 0 or 1,
and its interactions are renormalizable.

%%%%%
%%%%%%%%%%%%

A new vector boson, $W'$, of electric charge $+1$ would contribute only
to $C_A^\ell$. Such a boson must be associated with a new gauge symmetry,
which makes it difficult to allow large couplings to left-handed leptons.
One possibility is that $W$ and $W'$ mix, but the constraint from
electroweak data on mixing ($\lesssim 10^{-2}$) is too strong to allow
noticeable deviations in $D_s$ decays.
Another possibility is that some new vector-like fermions transform
under the new gauge symmetry and mix with the left-handed leptons.
Such mixing is also tightly constrained, especially by the
nonobservation of vector-like fermions at LEP and the Tevatron.
Overall, a $W'$ is inconsistent with Eq.~(\ref{eq:estimation-A}),
barring perhaps some finely-tuned elaborate model ({\it e.g.}, with
large $W$-$W'$ mixing whose electroweak effects are cancelled by other
particles).

A spin-0 particle of charge +1, $H^+$,
appears in models with two or more Higgs doublets.
Its interactions, in the 
mass eigenstate basis for charged fermions, include
\begin{equation}
H^+ \left( y_c \bar{c}_R s_L + y_s \bar{c}_L s_R 
+ y_\ell \bar{\nu}_\ell \ell \right) + {\rm H.c.},
\end{equation}
where $y_c,y_s,y_\ell$ are complex Yukawa couplings. 
The exchange of  $H^+$ induces $C_A^\ell =0$ and 
\begin{equation}
	C_P^\ell = \frac{1}{2}\left(y_c^* - y_s^*\right) y_\ell ~,
%  	C_P^\ell = (y_c^* - y_s^*) y_\ell/2 ~,
\end{equation}
taking $M$ equal to the $H^+$ mass.
If $H^+$ is the charged Higgs boson present in the Type-II
two-Higgs-doublet model, then $y_c/y_s = m_c/(m_s\tan^2\!\beta)$ so
that $C_P^\ell$ can have either sign
\cite{Hewett:1995aw}, but the Yukawa couplings are too
small to be compatible with Eq.~(\ref{eq:estimation-P}).
Other models lead to large constructive interference.
For example, a two-Higgs-doublet model where one doublet gives the 
$c$, $u$ (but not $d$, $s$, $b$, or $t$) and lepton masses, and 
has a vacuum expectation value of $\sim 2$ GeV, 
yields $|y_s| \ll y_\tau , y_c^* \sim O(1)$. Thus, $C_P^\ell > 0$ and
the limits (\ref{eq:estimation-P}) are satisfied for $M \lsim 500$ GeV. 
Furthermore, such a model explains why the deviations in 
$\tau\nu$ and $\mu\nu$ are comparable.
It is encouraging that this two-Higgs-doublet model does not induce 
tree-level flavor-changing neutral currents, and the off-diagonal 
couplings of $H^+$ are CKM suppressed. 
Given that this model has not been 
previously studied, its 1-loop contributions to flavor-changing processes 
(such as $b\!\to\! s\gamma$) need to be computed before deciding whether  
some fine tuning is required to evade experimental bounds.
 
The charge $-1/3$ and $+2/3$ exchanges correspond to leptoquarks.
A scalar charge $+2/3$ exchange arises for the  $(3,2,+7/6)$ set of 
$SU(3)_c\!\times\! SU(2)_W\!\times\! U(1)_Y$ charges.
This leptoquark appears, for example, in a new theory
of quark and lepton masses~\cite{Dobrescu:2008}.
Let $r = (r_u, r_d)$ be the doublet leptoquark, where $r_d$ is its
charge $+2/3$ component.
The interaction terms relevant here, written in the same basis
as~(\ref{eq:Leff4}), are 
$\lambda_{c\ell}r_d\bar{c}_R\nu_L^\ell+\lambda'_{s\ell}r_d\bar{s}_L\ell_R$.
The $r_d$ exchange gives $C_A^\ell=0$ and
$C_P^\ell=-\lambda_{c\ell}^*\lambda'_{s\ell}/4$.
%, for $M$ equal to the $r_d$ mass.
Since the leptoquark couplings can have any phase, the new amplitude 
can interfere constructively. 
Still, various flavor processes constrain the couplings of~$r$.
Even if its couplings to first-generation fermions were negligible, the
lepton-flavor violating decays $\tau\arrow\mu\bar{s}s$, where $\bar{s}s$
hadronizes to $\eta$, $\eta'$, $\phi$ or $K\bar{K}$, set a lower limit
on $M^2\!/|\lambda'_{s\tau}\lambda'_{s\mu}|$, which is hard to reconcile
with Eq.~(\ref{eq:estimation-P}).
One way out would be a model with two $r$ leptoquarks, with one 
coupling to $\tau$ and the other one to $\mu$.
The constraint from $\tau\arrow\mu\bar{s}s$ similarly disfavors spin-1
leptoquarks of charge~$+2/3$.

A scalar leptoquark of charge $-1/3$ (also discussed in
\cite{Dobrescu:2008}) arises in the case of two sets of 
$SU(3)_c\times SU(2)_W\times U(1)_Y$ charges: 
$(3,1,-1/3)$ or $(3,3,-1/3)$. Let us denote the former by $\tilde{d}$.
Its Yukawa couplings are given by 
\begin{equation}
	\tilde{d}\left[\kappa_\ell 
		\left(\bar{c}_L \ell_L^c - \bar{s}_L \nu_L^{\ell c} \right) +
		 \kappa^\prime_\ell \, \bar{c}_R \ell_R^c \right] +
	{\rm H.c.},
\end{equation}
where $\kappa_\ell$ and $\kappa^\prime_\ell$ are complex parameters.
These interactions are present, for example, in $R$-parity violating
supersymmetric models (their effect on $D_s\arrow e^+\nu$ 
has been analyzed in Ref.~\cite{Akeroyd:2002pi}).
The $\tilde{d}$ exchange, as in the last diagram of Fig.~\ref{fig:diagrams},
gives (for $M$ equal to the  $\tilde{d}$ mass)
\begin{equation}
	C_A^\ell = \frac{1}{4} \, |\kappa_{\ell}|^2 \;\;\;  , \;\;   \;     
	C_P^\ell = \frac{1}{4} \, \kappa_{\ell}\kappa^{\prime *}_{\ell}  ~~.
% 	C_A^\ell = |\kappa_{\ell}|^2/4 \;\;\;  , \;\;   \;     
% 	C_P^\ell =  \kappa_{\ell}\kappa^{\prime *}_{\ell}/4  ~~.
\end{equation}
For $|\kappa_{\ell}^\prime/\kappa_{\ell}| \ll m_\ell m_c/m_{D_s}^2$,
the  interference is automatically constructive [see Eq.~(\ref{eq:lq-rate})], 
and the resulting deviations in $\tau\nu$ and $\mu\nu$ are approximately equal
if $|\kappa_\mu| \approx |\kappa_\tau|$.
Moreover, there are no severe constraints from other processes on the couplings
$\kappa_{\ell}$ and $\kappa'_{\ell}$ with $\ell=\tau$ or $\mu$.
The~$\tilde{d}$ couplings to the electron can be forbidden by
a symmetry, and the ones to 
%its couplings to 
first-generation quarks could be small.
 
The $(3,3,-1/3)$ scalar leptoquark includes an $SU(2)_W$
component of charge $-4/3$ which mediates $\tau\arrow\mu \bar{s} s$.
The vector leptoquark of charge $-1/3$ %\cite{Yao:2006px} 
has the same problem.

\medskip
%%%%%%%%%%%%%%%%%%%%%%%%%%%%%%
{\it Conclusions.}---We have argued that the $3.8\sigma$ discrepancy 
between the standard model and the combined experimental measurements 
of $D_s\arrow\ell\nu$ appears so far to be robust, and thus 
it is worth interpreting it in terms of new physics. The upper bounds 
(\ref{eq:estimation-A}) and 
(\ref{eq:estimation-P}) on the scale of four-fermion operators are  
low enough to allow exploration of the underlying
physics at the LHC. 

A $\tilde{d}$ scalar leptoquark of charge $-1/3$ may solve the $D_s$ puzzle
without running into conflict with any other measurements.
At the LHC, the $\tilde{d}$ can be strongly produced in pairs,
and the final states would be $\ell^+\ell^-jj$, where $\ell$ is a 
$\tau$ or a $\mu$, and $j$ is a $c$-jet. Given that there are two 
$\ell j$ pairs, each of them forming a resonance at the $\tilde{d}$ mass, 
the backgrounds can be kept under control. The current limits on the 
 $\tilde{d}$ mass from similar searches at the Tevatron are around 200 GeV 
\cite{Abulencia:2005ua}.

An alternative explanation is provided by an $H^+$ exchange in a (new)
model where a Higgs doublet gives masses to the charged leptons and $c$
and $u$ quarks, and a second Higgs doublet gives masses to the down-type
and top quarks.
Both the leptoquark and charged Higgs solutions
lead naturally to comparable increases in the branching fractions for
$D_s\arrow\tau^+\nu$ and $D_s\arrow\mu^+\nu$, as suggested by the
data.

\smallskip
%%%%%%%%%%%%%%%%%%%%%%%%%%%%%%
{\it Acknowledgments.}---We thank 
%Patrick Fox, Enrico Lunghi, Sheldon Stone and Ruth Van de Water 
P.~Fox, E.~Lunghi, S.~Stone and R.~Van de Water 
for helpful discussions.
Fermilab is operated by Fermi Research Alliance, LLC, under 
%Contract DE-AC02-07CH11359 with the US Department of Energy.
US DoE Contract DE-AC02-07CH11359.

\end{document}